\documentclass[aps,twocolumn,nofootinbib]{revtex4}
\usepackage{graphicx}
\usepackage{dcolumn}
\usepackage{amssymb}
\usepackage{bm}
\usepackage{enumerate}

\usepackage{color}

\newcommand{\sla}[1]%
        {{\raise.15ex\hbox{$/$}\kern-.57em #1}}

\begin{document}
\title{Causal sets and conservation laws in tests of Lorentz symmetry}
\author{David Mattingly}
\affiliation{University of New Hampshire}
\email{davidmmattingly@comcast.net}

\begin{abstract}
Many of the most important astrophysical tests of Lorentz symmetry
also assume that energy-momentum of the observed particles is
exactly conserved.  In the causal set approach to quantum gravity a
particular kind of Lorentz symmetry holds but energy-momentum
conservation may be violated. We show that incorrectly assuming
exact conservation can give rise to a spurious signal of Lorentz
symmetry violation for a causal set. However, the size of this
spurious signal is much smaller than can be currently detected and
hence astrophysical Lorentz symmetry tests as currently performed
are safe from causal set induced violations of energy-momentum
conservation.
\end{abstract}


\maketitle

\section{Introduction}
The search for a complete and experimentally verified theory of
quantum quantum gravity is one of the most important open questions
in physics today.  Unfortunately, despite the efforts of numerous
eminent physicists, we do not yet have a theoretically complete
model for quantum gravity.  Since the natural energy scale of
quantum gravity is the Planck scale ($M_P= 10^{19}$ GeV) it is also
extremely difficult to perform direct experiments that support one
candidate model for quantum gravity over another.  Fortunately,
various ideas about quantum gravity have suggested that the defining
symmetry of special relativity, Lorentz symmetry, is not an exact
symmetry even at low energies. In string theory this can occur
perhaps by tensor VEV's~\cite{Kostelecky:1988zi} or noncritical
strings~\cite{Ellis:1999yd,Mavromatos:2007xe} while in loop quantum
gravity the ultimate fate of Lorentz symmetry and how it's
implemented is an open question
(c.f.~\cite{Gambini:1998it,Smolin:2005cz,Rovelli:2002vp,Bojowald:2004bb}).
Emergent spacetime models, for example analog
spacetime~\cite{Barcelo:2005fc}, contain Lorentz symmetry violation
(LV) at high energies.  Canonical noncommutative field theories also
contain Lorentz
violation~\cite{Carroll:2001ws,Anisimov:2001zc,Hinchliffe:2002km}
although they are invariant under the twisted Poincare
group~\cite{Chaichian:2004yh}. Note that there are also
noncommutative models where Lorentz invariance is deformed rather
than explicitly violated
(see~\cite{KowalskiGlikman:2004qa,Freidel:2007yu} and references
therein).

Of course any violation of Lorentz invariance must be very, very
small and therefore for any model with LV there is a severe
hierarchy problem. For example, in an effective field theory
context, mass dimension three operators that generate Lorentz
symmetry violation must be less than $10^{-32}$ GeV while the
dimension four operators are constrained at the level of
$10^{-28}$~\cite{Kostelecky:1999mr}. These extreme bounds mean that
any Lorentz violating theory must answer the question of why we have
such good approximate low energy Lorentz invariance.  One could fine
tune operators, but this is unnatural. In an attempt to avoid this,
many authors have looked at higher dimension operators where the
magnitude of the operator is suppressed by $M_P$ and so is
``hidden''. However, standard effective field theory arguments
require that the higher dimension operators dimensionally transmute
into lower dimension ones~\cite{Collins:2004bp}.  This means that
dimension five and six operators are constrained at the same order
as the dimension three and four operators and so LV alone is
extremely unlikely. It is possible to naturally suppress the
renormalizable operators by introducing new physics at scales just
above that currently accessible by particle detectors.  For example,
imposing supersymmetry~\cite{Bolokhov:2005cj} as well as Lorentz
violation stabilizes the generation of lower dimension operators at
the SUSY breaking scale.  If the SUSY breaking scale is given by
$m_{SUSY}$, and the original higher dimension operators are
suppressed by $M_P$, then SUSY will naturally suppress the dimension
three and four operators to the level of $m_{SUSY}^2/M_P$ and
$m_{SUSY}^2/M_P^2$ respectively.  For $m_{SUSY}<100$ TeV, this gives
a suppression to a level of approximately $10^{-9}$ GeV and
$10^{-28}$.  The dimension three operators are still in conflict
with observational bounds, but these can be eliminated by further
imposing CPT invariance.  This would make a model with low energy
SUSY and CPT invariance and LV sourced by Planck scale quantum
gravity viable, but requires significant new \textit{low energy}
physics as well (in this case SUSY and the assumption of CPT).

There is, of course, nothing a priori wrong with introducing new low
energy physics and SUSY certainly has a number of other compelling
features unrelated to LV.  If we restrict ourselves just to the
question of LV, however, then Occam's razor suggests that adding
\textit{two} new physical ideas is disfavored. Hence, instead of
either fine tuning or imposing new low energy physics in addition to
LV, it would be better perhaps to have a quantum gravity theory that
respects Lorentz invariance exactly. The discrete model we will
discuss for the rest of this paper, causal sets, is singled out
among discrete models as it is constructed to be Lorentz invariant
by definition. Recently, Sorkin, Bombelli, and
Henson~\cite{Bombelli:2006nm} proved that a causal set is Lorentz
invariant for an abstract operator that represents a measurement of
a preferred frame for a section of the causal set (the proof is that
such operators cannot exist).  In this work we argue that this
operator $D$ does not quite reflect the way that we currently
analyze many real Lorentz violating experiments and that such
experiments (in particular astrophysical tests) may theoretically
show ``spurious'' Lorentz violating effects if the underlying
spacetime is a causal set.  However, we also show that any effect is
much smaller than our current experimental sensitivities and can be
safely ignored.

The essence of our argument is that swerves, a hypothetical effect
on particle propagation in causal set theory~\cite{Dowker:2003hb},
can mimic a signal of LV in time of flight or threshold astrophysics
experiments where the expected flux of incoming high energy
particles is very low.  This may seem strange, as causal sets are
supposed to not give any LV signal.  However, swerves violate
energy-momentum conservation, which we usually assume holds exactly
in the vacuum when we analyze LV experiments.  This mismatch between
theory and assumption leads to a spurious LV signal.  Violations of
energy-momentum conservation are tightly constrained from cosmology
and the constraints are strong enough that this effect is irrelevant
at current sensitivities in LV experiments but may not be in the
future.  Furthermore, other ideas about quantum gravity have raised
the spectre of a small violation of energy-momentum conservation for
particles traveling in the vacuum(c.f
~\cite{Ellis:2000dy,Parentani:2007uq,AmelinoCamelia:2002ws}), so
exploring how this changes LV experiments is required if we are to
analyze the experiments properly.

An added benefit to this study is that, for a population of
particles, the violation of energy-momentum conservation predicted
from causal sets satisfies a low energy diffusion equation in
energy-momentum space. This equation is unique, and therefore our
result implies that we can rather generically neglect the effects of
any stochastic, Lorentz invariant violation of energy-momentum
conservation in Lorentz symmetry tests, whether or not we believe in
causal sets. Hence while causal sets are the motivation, the results
apply more broadly (although causal sets are the only model at this
point that hypothesizes such an effect). To put it differently,
assuming energy-momentum conservation is not a priori warranted when
searching for quantum gravity induced LV. However, the constraints
on Lorentz invariant conservation violation are so tight that we can
safely ignore it at our current level of sensitivity in LV tests,
independent of whatever the fundamental theory turns out to be.

Throughout the following discussion we choose metric signature $-2$
and units such that $\hbar=c=1$.

\section{Causal sets and the Poincare group}
Before we can discuss the fate of the Poincare group and
conservation laws in the context of causal sets, we need to know
what exactly a causal set is.  At its basic level, a causal set is a
partially ordered set (a poset), consisting of `points' $x,y,...$
and relations $x \prec y$ which encode causal ordering, i.e. $x
\prec y$ implies $x$ is to the past of $y$.  The ordering obeys two
other rules,  $x \prec y$, $y \prec z \Rightarrow x \prec z$ and $x
\not \prec x$. The first is simple transitivity such that if $x$ is
in the past of $y$ and $y$ is in the past of $z$, $x$ had better
also be in the past of $z$.   The other condition forbids closed
causal curves. Causal sets are usually considered to be finite, so
for any ordered pair $\{x,y\}$ with $x \prec y$, there are a finite
number of points $z$ that satisfy $x \prec z$ and $z \prec y$.
While it is certainly not necessary for a spacetime to be present
for a causal set to be defined, a useful picture is to think of a
causal set as a random lattice ``sprinkled'' in a Lorentzian
spacetime where there is an average of one point per every volume
$V_0$ (which will be assumed here to be a Planck volume $L_{Pl}^4$).
The causal set sprinkling therefore approximates $M$.  For more
information on causal sets, see~\cite{Henson:2006kf} and references
therein.

It has been argued that an \textit{individual} causal set is Lorentz
invariant~\cite{Bombelli:2006nm}, even locally, in a particular
sense: there is no experiment that can assign an intrinsic violation
of Lorentz invariance to a causal set at any point. Here ``intrinsic
violation'' means a frame that depends only on the sprinkling.  We
now briefly outline the proof in~\cite{Bombelli:2006nm}, restricting
ourselves to the simplest type of LV - a preferred frame. The
question is then, can a causal set define such a frame? Certainly a
regular discrete structure for spacetime does, but a causal set is a
random structure where the points are distributed in spacetime from
a Lorentz invariant probability distribution.  A preferred frame is
specified by the existence, in vacuum, of a unit future pointing
timelike vector field $u^a$ everywhere on the spacetime manifold
$M$.  Consider now a subset $\Omega$ of a causal set sprinkling that
approximates $M$. An experiment that assigns a LV preferred frame
can be represented as an operator $D$ that maps $\Omega$ to the unit
hyperboloid of future timelike vectors $H$, i.e. $u^a=D \circ
\Omega$.   $u^a$ transforms as an ordinary four vector under a
Lorentz transform $\Lambda$.  $D$ must not posses any intrinsic
preferred direction (to ensure that any LV comes from $\Omega$
itself) and therefore $D$ must commute with $\Lambda$ so that $D
\Lambda \Omega= \Lambda u^a$. The proof is then that no such
operator $D$ can exist and there is no experiment that can assign a
preferred frame to a local patch $\Omega$ of a causal set.  The
implication is that no LV experiment can ever assign a preferred
frame to a causal set intrinsically.

This result, on its face, is incompatible with another hypothesized
effect of causal sets, a violation of translation invariance due to
the so-called ``swerve'' effect~\cite{Dowker:2003hb}. The swerve
effect, which we discuss in more detail in the next section,
manifests itself at low energies via a Lorentz invariant diffusion
equation in momentum space.  An initial collection of particles with
some momentum distribution $\rho(p)$ therefore evolves into a
different state over (proper) time, which obviously violates
translation invariance.  The immediate question is then, how does
this not directly yield LV?  After all, translation violation allows
us to immediately define a frame by taking, for example, the
gradient of the evolving quantity as the operator $D$.  The answer
is that the frame so defined is not \textit{intrinsic} to the causal
set, but also involves the initial distribution $\rho(p)$. In
particular, if $\rho(p)$ is a Lorentz invariant function (which is
unrealistic physically but useful for this discussion) it is
preserved by the diffusion equation.  Hence there is neither a
violation of translation or Lorentz invariance in this case.  If,
instead, $\rho(p)$ is not Lorentz invariant then there can be
translation invariance violation, but both it and the corresponding
LV are functions of the causal set \textit{and} the initial LV
distribution.  There is a ``signal'' of LV, but it is not the
intrinsic LV which is forbidden by the proof outlined above because
our operator $D$ contains a preferred direction, which violates one
of the assumptions in the proof.

The difference between these two types of LV signal - a real signal
intrinsic to the underlying spacetime vs. a spurious signal due to
the combination of the preferred frame of an experiment with another
property of spacetime (in this case swerves) is a possibility which
has not arisen before in analysis of LV experiments.  Since all of
our observations and experiments that search for LV are not LI in
and of themselves (as they are local and happen in a particular
frame) we must consider it. Hence the question of LV in causal sets
for a practical experiment is not quite as straightforward as
presented in~\cite{Bombelli:2006nm}. The particular example we
examine in this paper is how the \textit{assumption} of translation
invariance, which is customary in LV experiments, can generate a
spurious LV signal from a causal set.

Translation invariance is a bad assumption due to the swerve effect,
as we a) single out a preferred Killing vector which is not an
intrinsic property of the causal set itself and b) neglect momentum
space diffusion due to swerves. The most common astrophysical tests
of LV are time of flight tests, where we compare the arrival time of
two high energy particles, and anomalous scattering/decay phenomena,
where we look for a modification to the scattering amplitude caused
by LV dispersion relations for the in/out free particle states.  As
will become clear below, when analyzing these phenomena we
\textit{must} assume something about translation invariance and
conservation of energy and momentum in order to put constraints on
any possible LV. Usually, translation invariance has been assumed to
hold exactly in flat space, as this is most consistent with a
straightforward field theoretical approach if the background LV
tensor fields are constant.\footnote{There are two major exceptions
to this, the doubly special relativity program
(see~\cite{KowalskiGlikman:2004qa} for an introduction) which
deforms both the translational and Lorentz subgroups of the Poincare
group, and spacetime foam ideas which couple LV to a fluctuating
dispersion term~\cite{Aloisio:2006nd,Aloisio:2002ed}.  These
approaches, however, have a significant a priori modification of
\textit{both} Lorentz invariance and translation invariance which
makes them distinct from the causal set approach.}  With causal
sets, this assumption isn't right and we must verify that a LV
signal is really due to LV and not the swerve effect.

\subsection{Swerves}
Consider the intuitive picture for swerves in~\cite{Dowker:2003hb},
that of a classical particle propagating on a random spacetime
lattice with mass $m$ and velocity $v$. The particle is constrained
to move from point to point, which poses a problem as generically
there is no probability of a lattice point lying on the future
worldline of the particle.  The particle must therefore ``swerve''
slightly so that it remains on the lattice and slightly change its
velocity.  A change in velocity is equivalent to the particle moving
to a different point on its mass shell, which is assumed to be
unchanged since causal sets are Lorentz invariant.   The net result
of swerving is that a collection of particles initially with an
energy-momentum distribution $\rho(p)$ will diffuse in momentum
space along their mass shell according to the unique Lorentz
invariant diffusion equation~\cite{Dowker:2003hb,Sorkin:1986nm},
\begin{equation} \label{eq:diffusion}
\frac {\partial \rho} {\partial \tau} = k \nabla^2_P \, \rho -
m^{-1} p^\mu \partial_\mu \rho .
\end{equation}
Here $k$ is the diffusion constant, $\nabla^2_P$ is the Laplacian in
momentum space on the mass shell of the particle,  $\tau$ is the
proper time, $\partial_\mu$ is an ordinary spacetime derivative, and
$m$ is the mass.  While the underlying classical picture is almost
certainly incorrect,(\ref{eq:diffusion}) is the \textit{unique}
Lorentz invariant diffusion equation.  Therefore if there is any
type of stochastic violation of Poincare invariance due to a random
discrete structure underlying spacetime a la causal sets, at lowest
order it should be described by (\ref{eq:diffusion}).  There is a
very tight limit on $k$ for neutrinos from
cosmology~\cite{Kaloper:2006pj}, $k<10^{-61} \mathrm{GeV^3}$, which
comes from limits on the amount relic neutrinos can contribute to
hot dark matter (and hence how much energy they can gain from
swerves).  While theoretically each particle species could have a
different $k$, it would be rather unnatural if they were too far
apart, especially as the source of the diffusion is supposed the
same discrete spacetime structure.  Therefore we shall take
$k<10^{-61} \mathrm{GeV^3}$ to be our rough constraint for all
particles.  As it turns out, any value of $k$ even close to
$10^{-61} \mathrm{GeV^3}$ makes energy-momentum conservation
violation irrelevant for LV searches, so this is a perfectly safe
assumption.

Energy is bounded below and so a particle initially at rest will
increase its kinetic energy due to swerves.  Since $k$ is so low, we
can make a simplifying assumption for any collection of particles
that are not of cosmological age.  In the initial rest frame of the
particle the swerves can first be treated in the non-relativistic
limit for a certain period of time that depends on $k$.  This is
obvious as over time energy is being added to the particle via
swerves, but as long as the total energy is less than the rest mass
the appropriate limit is the non-relativistic one.  In the
non-relativistic limit, (\ref{eq:diffusion}) simplifies to
\begin{equation} \label{eq:diffusiont}
\frac{\partial \rho} {\partial t} = k \nabla^2_P  \rho
\end{equation}
where $t$ is now the coordinate time and $\nabla^2_P$ is the
standard Laplacian operator on momentum space $\mathbb{R}^3$. The
solution for a collection of particles all initially at rest has
been derived in~\cite{Kaloper:2006pj} and is given by
\begin{equation}
\rho(p)=(4\pi k t)^{-3/2} e^{- \frac {p^2} {4kt}}
\end{equation}
which is the thermal distribution for a non-relativistic gas at
temperature $T=2kt/m$.

We can now ask how long a collection of initially cold particles
must exist for the non-relativistic approximation to break down.
This happens when the temperature is roughly equal to the mass,
which gives $t\approx m^2/(2k)$.  For a neutrino of mass $10^{-1}$
eV, the approximation breaks down after $10^{17}$ seconds, while for
electrons and protons the approximation is always good as the
breakdown time is far longer than the age of the universe.  After a
population becomes relativistic it will still gain energy but not as
quickly since the proper time is shorter.  Therefore we know that
the energy gained (per neutrino) by a population of initially cold
neutrinos over the lifetime of the universe (also approximately
$10^{17}$ seconds) is no more than about the rest mass of the
neutrino and the energy gain for other species is far less. This
neglects the effects of cooling due to expansion, etc. which are
dealt with in~\cite{Kaloper:2006pj}, however we can ignore these
secondary effects for our purposes.  Now consider a population of
particles with a high gamma factor $\gamma$ that have traveled to
earth from a source at one Gpc.  The lifetime of the particle in our
frame $O$ is $10^9 \rm{years}=10^{16} \rm{seconds}$.  However, the
time in the (initial) rest frame $O'$ of the particle is much
shorter, $10^{16}/ \gamma$ seconds.   Since any particle can gain no
more than $m_\nu$ in energy over $10^{17}$ seconds and the
propagation time in $O'$ is much shorter, very little energy is
gained in $O'$ due to swerves. Hence in $O'$ the particles are all
still very non-relativistic once we include the swerve effect as
long as $\gamma \gg 1$ and the non-relativistic approximation is
valid for their entire lifetime.

Of particular interest is the average deviation from the initial
energy $E_i$ of a particle. If we define $\Delta
E/E_i=|(E_f-E_i)|/E_i$, where $E_f$ is the energy at time of
measurement, then from the discussion above we know that $\Delta
E/E_i$ is at least less than $\gamma^{-1}$ (on average) for any
species. We can see this more explicitly by noting from above that
for a neutrino with a lifetime of the age of the universe at rest in
our frame, the total energy gain is at best the mass of the
neutrino. If instead the neutrino is boosted with respect to our
frame, the proper time is reduced by a factor of $\gamma^{-1}$ and
so the total energy gain is reduced also by $\gamma^{-1}$ from its
initial energy $E_i=\gamma m$.  Therefore for neutrinos the ratio
$\Delta E/E_i \leq \gamma^{-1}$. For TeV and above astrophysical
neutrinos, which we are primarily interested in, $\Delta E/E_i \leq
\gamma^{-1} = 10^{-13}$ at worst. The swerve rate is slower for
other species, so this bound holds for them as well. This gain is
very small, however LV tests can be sensitive to fractional changes
in energy of order $10^{-28}$ so it is not a priori obvious that
swerves are irrelevant.  We now turn to how this diffusion affects
these LV tests.

\section{Swerves and astrophysical tests of LV}
\subsection{Time of flight}
Time of flight tests are perhaps the simplest type of LV experiment.
In these experiments one looks for delays in the arrival time of
high energy particles from distant astrophysical events.   A time of
flight experiment compares the arrival time of at least two high
energy particles and the time delay between arrivals can be caused
by three distinct effects.   The first is source effects - the
particles are not produced at the same time or location in the
astrophysical event.  The source delay can be quite long, in the
case of neutrinos from gamma ray bursts the delay time of a neutrino
associated with the burst can be days. The second type of delay is
detector response.  These delays are usually small and known and we
will not consider them further.  Finally, an unexplained time delay
is usually considered evidence that the speed of propagation of the
high energy particles is different than the speed of light.   This
is the ``interesting'' signal of a violation of Lorentz invariance.

\subsubsection{Time of flight in field theory}

It will be useful to discuss in detail how these experiments work in
a very concrete and established framework first, before we consider
the causal set scenario, so let us analyze time of flight in a field
theory context first.  LV occurs when fields are coupled in vacuum
to a non-zero tensor field other than the metric.  If there exists a
preferred frame in nature, the fields couple to a unit timelike
vector $u^a$ which describes the preferred frame.  There are many
ways a field could couple in both the matter and gravitational
sectors~\cite{Colladay:1998fq,Jacobson:2000xp,Myers:2003fd,Bolokhov:2007yc},
for a review of both the renormalizable and non-renormalizable
operators see~\cite{Mattingly:2005re}.  The free field mass
dimension five couplings and below are already tightly constrained,
so we consider here an unconstrained operator - that of a dimension
six CPT even operator. While terms like this may be able to be
tested in ultrahigh energy neutrino observatories in the future and
hence are intrinsically relevant, we choose this term for another
reason:  if the swerve effect is irrelevant for tests of this term
it is certainly irrelevant for any LV time of flight test we could
conceivably perform in the near future.  The dimension six CPT even
LV modification to the kinetic term for a fermion is
\begin{equation} \label{eq:actionfermion}
\mathcal{L}_f=\overline {\psi} (i\sla{\partial} - m) \psi -
    \frac {i} {E_{Pl}^2} \overline {\psi} (u
\cdot \partial)^3 (u \cdot \gamma) (\alpha_L P_L + \alpha_R P_R)
 \psi
\end{equation}
where $P_{R,L}$ are the usual right and left chiral projection
operators and $\alpha_{R,L}$ are coefficients.  Usually
$\alpha_{R,L}$ are assumed to be O(1).  We choose the operator to be
suppressed by the Planck scale $E_{Pl}$ as this would be the natural
scale if the term was generated by some theory of quantum gravity at
$E_{Pl}$.

The Hamiltonian corresponding to ({\ref{eq:actionfermion}) commutes
with the helicity operator, hence the eigenspinors of the modified
Dirac equation will also be helicity eigenspinors.  We now solve the
free field equations for the positive frequency eigenspinor $\psi$.
Assume the eigenspinor is of the form $\psi_s e^{-i p \cdot x}$
where $\psi_s$ is a constant four spinor and $s=\pm1$ denotes
positive and negative helicity. Then the Dirac equation becomes the
matrix equation
\begin{equation} \label{eq:fermiondirac}
\left(%
\begin{array}{cc}
  -m  & E-sp - \alpha^{(6)}_R \frac {E^3} {E_{Pl}^2}  \\
E+sp  - \alpha^{(6)}_L \frac {E^3} {E_{Pl}^2}   & -m  \\
\end{array}%
\right) \psi_s=0
\end{equation}
The dispersion relation, given by the determinant of
(\ref{eq:fermiondirac}), is
\begin{eqnarray} \label{eq:dispfermion} \nonumber
E^2 -( \alpha^{(6)}_RE^3)
(E+sp)\\
-(\alpha^{(6)}_LE^3) (E-sp)=p^2+m^2
\end{eqnarray}
where we have dropped terms quadratic in $\alpha^{(6)}_{R,L}$ as
they are small relative to the first order corrections for those
terms.

At $E\gg m$, as appropriate for high energy astrophysical particles,
the helicity states are almost chiral, with mixing due solely to the
particle mass. Note also that at energies $E>>m$, we can replace $E$
by $p$ at lowest order, which yields the approximate dispersion
relation
\begin{equation} \label{eq:dispfermionhighE}
E^2 =p^2+m^2  +2\alpha_{R,L} \frac{p^4} {E_{Pl}^2}.
\end{equation}
Positive coefficients correspond to superluminal propagation, i.e.
$v=\partial E/\partial p
> 1$, while negative coefficients give subluminal propagation. In either case, the speed of astrophysical particles does not asymptote to $c$ as the energy increases.

The group velocity $\partial E/\partial p$ is
\begin{equation}
v=1-\frac {m^2} {2p^2} + 3\alpha_{R,L} \frac {p^2} {E_{Pl}^2} = 1 +
\Delta v.
\end{equation}
For a source at distance $d$ from earth the arrival time $t_a$ of a
high energy particle is $t_a=d/v$.  The difference $\Delta t_{LV}$
between a light pulse emitted from the source at the same time as
the particle is
\begin{eqnarray} \label{eq:timedelay}
\Delta t_{LV}= t_{light}- t_a= d-\frac {d} {v} = d(\frac {v-1} {v})
\approx d \Delta v \\ \nonumber = d\big{(} -\frac {m^2} {2E^2} +
3\alpha_{R,L} \frac {E^2} {E_{Pl}^2} \big{)}
\end{eqnarray}
where we have replaced $p$ by $E$ in the high energy limit.

The $\alpha_{R,L}$ term in (\ref{eq:timedelay}) grows with energy.
In order to establish the best possible constraints we therefore
need to look at the highest energy particles.  The best chance we
have in a time of flight experiment to see LV at this order is in
the comparison of the arrival time of high energy neutrinos produced
by GRB's with the prompt emission arrival.  The flux at these
energies is very low which has both positive and negative
ramifications.  On the negative side we require large detectors like
ICECUBE to see any appreciable flux of ultra high energy GRB
neutrinos.  However the background flux is also very low.  Recently,
it has been proposed by Jacob and Piran~\cite{Jacob:2006gn} that
since the background is so low even the detection of a single
neutrino event days or weeks after a GRB can be associated with the
GRB and used to bound $\alpha_{R,L}$ in a time of flight experiment.
While this approach has other problems, primarily long source
delays~\cite{GonzalezGarcia:2006na} due to the GRB fireball
mechanics, it raises an interesting question with regards to causal
sets - what happens in the swerve picture when there are very, very
low statistics?

\subsubsection{Time of flight with swerves}
In time of flight experiments with large fluxes, swerves aren't a
problem.  For a strong multiparticle signal the \textit{average}
arrival time is still that as predicted by special relativity.
However, for a single particle signal with measured energy $E$ there
is no concept of averaging and the arrival time will not be that
predicted by special relativity.  The reason is simple - during
propagation the particle's energy is not $E$ and the particle's
velocity is not $v=1-m^2/(2E^2)$.  Therefore to conclusively ascribe
a time delay to a LV dispersion as in (\ref{eq:timedelay}) we need
to make sure the same delay cannot be due to swerves.

We can overestimate the time of flight delay for a typical particle
compared to special relativity by considering a particle propagating
with energy $E_f+\Delta E$, where $E_f$ is the measured (final)
energy and $\Delta E$ is the average deviation for a single particle
introduced previously.  It is an overestimate since we apply the
energy difference over the entire propagation of the particle.  The
dispersion relation is simply the relativistic dispersion relation,
so $\Delta t_S$ is
\begin{equation}
\Delta t_S = d\left( -\frac {m^2} {2(E_f+ \Delta E)^2}
\right)=d\left( - \frac {m^2} {2E_f^2(1+\Delta E/E_f)^2} \right)
\end{equation}
$\Delta E/E_f < \gamma^{-1} \ll 1$ and we therefore have
\begin{equation} \label{eq:deltatswerves}
\Delta t_S=d\left( - \frac {m^2} {2E_f^2} (1-2 \frac {\Delta E}
{E_f}) \right).
\end{equation}
Comparing (\ref{eq:deltatswerves}) with (\ref{eq:timedelay}) we see
that an experiment sensitive to LV at our chosen order is also
sensitive to swerves if
\begin{equation}
3\alpha_{R,L} \frac {E_f^2} {E_{Pl}^2} \approx \frac {m^2} {E_f^2}
\frac {\Delta E} {E_f}.
\end{equation}
Rewriting this equation for $\Delta E/E_f$ we have
\begin{equation} \label{eq:swervevsLV}
 \frac {\Delta E} {E_f}= 3\alpha_{R,L} \frac {E_f^4} {m^2 E_{Pl}^2}.
\end{equation}

There is no intrinsic size to the LV term in (\ref{eq:swervevsLV})
and depending on how accurate a LV time of flight measurement is, it
could theoretically also probe swerves.  However, the duration of a
long GRB can be of order 1000 seconds and it is unclear when in the
burst emission the neutrinos will occur.  Therefore there is an
intrinsic unknown source delay of at least 1000 seconds\footnote{The
actual value is of course dependent on the exact mechanics of the
GRB and can be shorter.  We use this value for illustrative purposes
as an order of magnitude estimate.}.  Any significant time delay
must be greater than this value and hence there is a lower limit on
a meaningful $\Delta t$ (swerve or LV induced).  For a GRB with this
value of $\Delta t$ and a distance of 1 Gpc, $\Delta t/d \approx
10^{-14}$ seconds.  With (\ref{eq:timedelay}) this establishes a
lower limit that the LV dispersion term of
\begin{equation}
|\alpha_{R,L} \frac {E_f^2} { E_{Pl}^2}| \geq 10^{-14}
\end{equation}
which must be satisfied if we are to see anything meaningful in a
time of flight GRB experiment.  If we are conservative and take our
high energy neutrinos to be above 1 TeV (actual proposed energies
are much higher) and a neutrino mass of approximately 0.1 eV, the
gamma factor is $10^{13}$. From (\ref{eq:swervevsLV}), $\Delta
E/E_f$ from swerves must then be greater than $10^{12}$, which is 25
orders of magnitude larger than the upper limit for swerving
neutrinos.  Hence swerves are completely and totally irrelevant.
This result can easily be generalized to other forms of LV
dispersion and in none are astrophysical time of flight experiments
sensitive to swerves by many orders of magnitude.

\subsection{Anomalous particle interactions}
Anomalous particle interactions are much more sensitive tests of LV
than time of flight tests and also of course require assumptions
about energy-momentum conservation.  Therefore they will also be
sensitive to the swerve effect at some level.  There are two types
of anomalous interactions.  The first type is when the interaction
occurs only in the LV model. The second type is when the interaction
begins to occur at a certain energy and this energy is different for
Lorentz invariant versus LV models.  We deal with each type of
interaction separately as the effect of swerves is
different.\footnote{We are not considering particle creation due to
the time dependence of the spectrum of an initial flux of particles
due to swerves or how to correctly define initial/final states in a
model without asymptotic translation invariance.  These questions
must also be answered in regards to the swerve effect but are
outside the scope of this paper.}

\subsubsection{New particle interactions}
Consider a proton with dispersion relation like that given in
(\ref{eq:dispfermionhighE}).  If $\alpha_{R,L}>0$ then at high
energies protons become unstable and emit photons in what is known
as the ``vacuum Cerenkov effect''.  They rapidly lose energy via
this process and hence the existence of ultra high energy cosmic ray
protons (for which this process cannot be occurring) limits how
positive $\alpha_{R,L}$ can be.  This is an example of a test that
uses a particle interaction completely absent in usual Lorentz
invariant physics to limit possible LV dispersion relations.  The
key to these tests is the energy-momentum conservation equations,
which tell us how much parameter space is available for the reaction
- zero in the Lorentz invariant case and non-zero with LV.  Naively
then, it seems reasonable that a fluctuating energy-momentum of the
initial and/or final states may allow for new reactions to also
occur.  We now show that in the case of causal sets, this is not
true.  If we consider just a swerve induced change in the
energy-momentum of the initial and final states then if a reaction
does not occur in straightforward special relativity it does not
occur in causal sets.  Note that in the following discussion we have
implicitly assumed that the swerve effect for massless particles
yields diffusion in a null cone in energy-momentum space (the $m
\rightarrow 0$ limit of a mass shell).

The idea is very simple.  Let us consider a generic multi-particle
reaction $i_1 + i_2 + ... \rightarrow f_1 + f_2 +...$ where $i_a$
are the incoming particles and $f_b$ are the outgoing particles.  If
the reaction does not occur in Lorentz invariant physics it means
that there is no solution to the conservation equation
\begin{equation}
p^\mu_1 + p^\mu_2 + ...=q^\mu_1 + q^\mu_2 + ...
\end{equation}
where $p_a^\mu$ (the incoming 4-momenta) and $q_b^\mu$ (the outgoing
4-momenta) are subject to the on-shell constraints $p_a^2=m_a^2,
q_b^2=m_b^2$.  In a LV theory the on-shell constraints change, in
causal sets they do not.  In a causal set, the incoming and outgoing
momenta are, however, modified by a fluctuation term $\Delta p_a,
\Delta q_b$ as the particle will swerve during the course of the
interaction.  We therefore rewrite the conservation equation in the
causal set approach as

\begin{equation}
p^\mu_1 + \Delta p^\mu_1 + p^\mu_2 + \Delta p^\mu_2 + ...=q^\mu_1 +
\Delta q^\mu_1 + q^\mu_2  + \Delta q^\mu_2...
\end{equation}
subject to the on-shell constraints $(p_a + \Delta p_a)^2=m_a^2,
(q_b + \Delta q_b)^2=m_b^2$ since the fluctuations must keep all
particles on-shell.  This though is just a relabeling of momenta and
doesn't change any physics - we can define new momenta
$\bar{p}^\mu_a = p^\mu_a + \Delta p^\mu_a$ etc. and the conservation
and constraint equations take the exact same form as before.
Therefore there is still no solution and the reaction doesn't happen
with swerves either.

There is a caveat to the above argument.  In keeping with other LV
threshold tests, where one looks at only initial and final states,
we have not considered the effect of swerves in the interaction
region. This is more dangerous for causal sets, as field theory on
causal sets is not well developed enough to know what the swerve
effect might do to virtual states that are not necessarily on-shell.
Hence this conclusion can not be considered absolutely concrete
until we understand more of quantum field theory on a causal set.
However, since no LV reactions have been seen to date it is likely
that causal set QFT does respect LI to a very good approximation
even when quantum effects are taken into account.

\subsubsection{Modification of existing energy thresholds}
The situation is different for reactions where there is a Lorentz
invariant solution.  Here, the swerve effect can change the energy
the reaction occurs at, although the shift is tiny.  Let us take a
specific reaction, pion production by proton-photon scattering, $p +
\gamma \rightarrow p + \pi^0$.  This reaction is important in LV
tests as the high energy cosmic ray spectrum should exhibit a cutoff
(the Greisen-Zapsetin-Kuzmin cutoff) around $5 \times 10^{19}$ eV
due to the scattering of cosmic ray protons off the cosmic microwave
background.  LV tests can shift this cutoff, and the recent
confirmation of the GZK cutoff by HiRes~\cite{Abbasi:2007sv} and the
Pierre Auger observatory~\cite{Yamamoto:2007xj} constrains some LV
models.  Again, we have a limited number of events (although with
Auger the statistics are getting rapidly better) and so one might
wonder if the random nature of swerves can cause a spurious signal.
Theoretically this is true, but  the effect of swerves on GZK
protons is negligible, even if we wildly overestimate the length of
time swerves have to act.  A GZK proton has a gamma factor of near
$10^{10}$, which means that its maximum proper lifetime is
$10^{17}/\gamma=10^7$ seconds if it was generated very early in the
universe.  The maximum amount of energy the proton can gain in its
initial rest frame, using our constraint for $k$, is $10^{-30}$ GeV,
which in turn implies that the shift in energy possible for a GZK
proton in our frame is $10^{-20}$ GeV.  The GZK cutoff will hence be
broadened by at best the insignificant amount of $10^{-20}$ GeV and
therefore swerves are irrelevant in the GZK reaction.  This type of
analysis can be done for many different threshold reactions, for
example the scattering of high energy TeV photons off the infrared
background.  In all cases the limits on $k$ are strong enough that
swerves have no appreciable effect.

\section{Conclusion}
LV searches in astrophysics rely not only on assumptions about the
nature of the LV model being explored, but also on energy-momentum
conservation of the particles involved.  In this paper we have
argued that causal sets, while intrinsically Lorentz invariant, can
technically introduce spurious LV signals if we deal with practical
experiments and make the usual assumption that energy-momentum is
conserved due to the swerve effect.  However, we have also shown
that existing limits on swerves imply that any errors introduced are
negligible at current experimental sensitivity.  A bonus is that the
diffusion equation (\ref{eq:diffusion}) that describes the swerve
effect is the unique low energy Lorentz invariant diffusion
equation.  Any statistical process that respects Lorentz invariance
but causes fluctuations in energy-momentum should therefore be
described by this equation and we can rather generically conclude
that any signal of LV in a time of flight or anomalous particle
interaction experiment cannot be due instead to a Lorentz invariant
modification of translation invariance.

\section{Acknowledgements}
We thank Stefano Liberati for useful comments on a draft of this
paper.

  \end{document}